\documentclass[useAMS,usedcolumn,usenatbib,usegraphicx]{mn2e}

\usepackage{epsfig,graphicx,natbib}
\usepackage{./reference/mycite}
\usepackage{amssymb}
\usepackage{amsfonts}
\usepackage{amsmath}
\usepackage{color}
 
\begin{document}

\title[Galaxy Zoo:Morphology via Machine Learning]{Galaxy Zoo: Reproducing Galaxy Morphologies via Machine Learning \thanks{This publication has been made possible by the participation of more than 100,000 volunteers in the Galaxy Zoo project. Their contributions are individually acknowledged at http://www.galaxyzoo.org/Volunteers.aspx}} 
\author[M. Banerji et al.]{ \parbox{\textwidth}{
Manda Banerji$^{1,2}$, \thanks{E-mail: mbanerji@ast.cam.ac.uk},
Ofer Lahav$^{1}$, 
Chris J. Lintott$^{3}$,
Filipe B. Abdalla$^{1}$,
Kevin Schawinski$^{4,5,6}$,
Steven P. Bamford$^{7}$, 
Dan Andreescu$^{8}$,
Phil Murray$^{9}$,
M. Jordan Raddick$^{10}$, 
Anze Slosar$^{11}$, 
Alex Szalay$^{10}$,
Daniel Thomas$^{12}$, 
Jan Vandenberg$^{10}$ \\
}
  \vspace*{6pt} \\
$^{1}$Department of Physics and Astronomy, University College London,
Gower Street, London, WC1E 6BT, UK.\\
$^{2}$Institute of Astronomy, University of Cambridge, Madingley Road, Cambridge CB3 0HA, UK.\\
$^{3}$Department of Physics, Denys Wilkinson Building, Keble Road, Oxford OX1 3RH, UK.\\
$^{4}$Department of Physics, Yale University, New Haven, CT 06511, USA \\
$^5$Yale Center for Astronomy \& Astrophysics, Yale University, P.O. Box 208121, New Haven, CT 06520, USA \\
$^6$Einstein Fellow \\
$^7$Centre for Astronomy and Particle Theory, School of Physics \& Astronomy, University of Nottingham, \\ University Park, Nottingham, NG7 2RD, UK. \\
$^8$LinkLab, 4506 Graystone Avenue, Bronx, NY 10471, USA. \\
$^9$Fingerprint Digital Media, 9 Victoria Close, Newtownards, Co.Down, Northern Ireland, BT23 7GY \\
$^{10}$Department of Physics and Astronomy, Johns Hopkins University, 3400 N. Charles Street, Baltimore, MD 21218, USA. \\
$^{11}$Berkeley Center for Cosmological Physics, Lawrence Berkeley National Laboratory \& Physics Department,\\ University of California, Berkeley, CA 94720, USA.\\
$^{12}$Institute of Cosmology and Gravitation, University of Portsmouth, Mercantile House, \\ Hampshire Terrace, Portsmouth, Hants PO1 2EG, UK
}

\maketitle

\begin{abstract}

We present morphological classifications obtained using machine learning for objects in SDSS DR6 that have been classified by Galaxy Zoo into three classes, namely early types, spirals and point sources/artifacts. An artificial neural network is trained on a subset of objects classified by the human eye and we test whether the machine learning algorithm can reproduce the human classifications for the rest of the sample. We find that the success of the neural network in matching the human classifications depends crucially on the set of input parameters chosen for the machine-learning algorithm. The colours and parameters associated with profile-fitting are reasonable in separating the objects into three classes. However, these results are considerably improved when adding adaptive shape parameters as well as concentration and texture. The adaptive moments, concentration and texture parameters alone cannot distinguish between early type galaxies and the point sources/artifacts. Using a set of twelve parameters, the neural network is able to reproduce the human classifications to better than 90\% for all three morphological classes. We find that using a training set that is incomplete in magnitude does not degrade our results given our particular choice of the input parameters to the network. We conclude that it is promising to use machine-learning algorithms to perform morphological classification for the next generation of wide-field imaging surveys and that the Galaxy Zoo catalogue provides an invaluable training set for such purposes.  

\end{abstract}


\section{Introduction}

Classification of galaxies has been a long-term goal in astronomy (e.g. \citet{Vandenberg:98} and references therein). While classification by human eye is still common, there have been several attempts to use machine learning techniques. For example, \citet{Lahav:95, Lahav:96} showed that artificial neural networks can successfully reproduce visual classifications. In recent years, artificial neural networks have gained prominence as a succesful tool for calculating photometric redshifts e.g. \citep{Collister:ANNz, Firth:nn} particularly with regard to the next generation of galaxy surveys e.g. \citep{Banerji:DES, Abdalla:DUNE}. However, artificial neural networks were first applied to astronomical data sets in order to classify stellar spectra \citep{vonHippel:94, Bailer:98} and galaxy morphologies \citep{MJL:92, Naim:95, Folkes:96}. Astronomical data sets have grown considerably in size in the last decade owing largely to the advent of mosaic CCDs that can be used on large telescopes in order to image large areas of the sky down to very faint magnitudes. The Sloan Digital Sky Survey (SDSS) \citep{York:SDSS} has led to the construction of a data set of around 230 million celestial objects. This data set has already been used for morphological classification using automated machine learning techniques \citep{Ball:04}. Future generations of wide-field imaging surveys such as the Dark Energy Survey\footnote{https://www.darkenergysurvey.org}, PanStarrs\footnote{http://pan-starrs.ifa.hawaii.edu/public/} and LSST\footnote{http://www.lsst.org} will reach new limits in terms of the size of astronomical data sets. Clearly, automated classification algorithms will prove invaluable for the analysis of such data sets but these algorithms are yet to be applied on such scales. 

The Galaxy Zoo project\footnote{http://www.galaxyzoo.org} launched in 2007 has led to morphological classification of nearly 1 million objects from the SDSS DR6 through visual inspection by more than 100,000 users \citep{Lintott:zoo}. This project has resulted in a remarkable data set that can be used for studies of the formation and subsequent evolution of galaxies in our Universe. In \citet{Lintott:zoo}, the Galaxy Zoo classifications have been compared to those by professional astronomers showing that there is remarkable agreement between classifications by members of the general public and the professionals. The biases in the Galaxy Zoo classifications have been studied in detail by \citet{Bamford:08} who go on to correct for these biases and use the dataset to study the relationship between galaxy morphology, colour, environment and stellar mass. This data set however also presents us with the unique opportunity to compare human classifications to those from automated machine learning algorithms, on an unprecedented scale. If the neural network is shown to be as successful as humans in separating astronomical objects into different morphological classes, this could save considerable time and effort for future surveys while ensuring uniformity in the classifications. 

In this paper we explore the ability of artificial neural networks to classify astronomical objects from the SDSS into three morphological types - early types, spirals and point sources/artifacts. In $\S$ \ref{sec:gz} we describe the Galaxy Zoo catalogue. In $\S$ \ref{sec:ann}, the artificial neural network method is presented. $\S$ \ref{sec:params} details the different choices of input parameters that are used for classification. We present our results in $\S$ \ref{sec:results} and draw some conclusions in $\S$ \ref{sec:conclusion}.

\section{The Galaxy Zoo Catalogue}

\label{sec:gz}

Galaxy Zoo is a web-based project that aimed to obtain morphological classifications for roughly a million objects in the Sloan Digital Sky Survey by harnessing the power of the internet and recruiting members of the public to perform these classifications by eye. The first part of this project is now complete and the morphological classifications subsequently obtained have been described in detail in \citet{Lintott:zoo} where these classifications have also been shown to be credible based on comparison with classifications by professional astronomers. The classifications have also been used in a number of interesting science papers e.g. in the identification of a sample of blue early type galaxies in the nearby Universe \citep{Schawinski:zoo} and to study the spin statistics of spiral galaxies \citep{Land:spin} and the power of this data set is proving enormous for studies of both galaxy formation and evolution \citep{Bamford:08}. Our goal is now to assess whether morphological classifications such as those from Galaxy Zoo can be reproduced for even larger data sets likely to become available with the next generation of galaxy surveys through the use of automated machine learning algorithms such as artificial neural networks. Before we proceed, we caution the readers that the Galaxy Zoo catalogue is not represented by a simple selection function as is the case for both volume and flux-limited samples. It contains objects from both the Main Galaxy Sample (MGS) as well as the Luminous Red Galaxy (LRG) sample of the SDSS and as such over represents the number of distant red galaxies in the Universe. This is not particularly important for the aims of this paper as we are simply attempting to reproduce the human classifications using machine learning. However, when using these classifications for scientific analysis, further cuts could be applied in order to remove this bias. 

The Galaxy Zoo catalogue that we use in this paper is the combined weighted sample of \citet{Lintott:zoo}. This contains morphological classifications for 893,212 objects into four morphological classes - ellipticals, spirals, mergers and point sources/artifacts. Note that the only stars in the Galaxy Zoo catalogue are bright stars with halos or diffraction spikes which are thus interpreted as extended objects by the automatic star/galaxy separation criteria and the only point sources included are those objects whose spectra are best fit by galaxy templates. The classification by each user on each object is weighted such that users who tend to agree with the majority are given a higher weight than those who don't. The final morphology of the object is then a weighted mean of the classifications of all users who analysed it. Full details of the weighting scheme are provided in \citet{Lintott:zoo}. The weighted catalogue of objects contains the SDSS object ID and three additional columns with the weighted fraction of vote of the galaxy being an elliptical, spiral or point source/artifact - i.e. classified as \textit{don't know} by Galaxy Zoo users - between 0 and 1. Any star-forming irregular galaxies are also put into this \textit{don't know} class by the Galaxy Zoo users. If the sum of the fractional votes in each of these three classes is less than 1, the remaining fraction of vote is assigned to the merger class. Note that this data set is affected by a luminosity, size and redshift dependent classification bias as is the case for most morphologies derived from flux-limited data sets. \citet{Bamford:08} have derived corrections to remove this classification bias from the data. However, in this paper we work with the original catalogue of morphologies as the classification biases are not particularly important for the aims of this paper. 

We match the Galaxy Zoo catalogue to the SDSS DR7 PhotoObjAll\footnote{Note that the SDSS object IDs correspond for objects in DR6 and DR7} catalogue in order to obtain input parameters for the neural network code. These input parameters are described in detail in $\S$ \ref{sec:par}. Before input into the neural network, we apply cuts on our sample and remove objects that are not detected in the $g$, $r$ and $i$ bands and those that have spurious values and large errors for some of the other parameters used in this study. \citet{Darg:09} have already discussed issues to do with merger classification within Galaxy Zoo and constructed a sample of $\sim$3000 merging pairs from the Galaxy Zoo data. In this paper, we classify objects as ellipticals, spirals or point sources/artifacts using our machine learning code and note that the \citet{Darg:09} data set may be used in future for the classification of mergers although this has not been attempted in this paper. We therefore also remove the few well classified mergers with a fraction of vote of being a merger greater than 0.8 from the sample as we are not attempting to classify the mergers in this work. This leads to a sample of $\sim$800,000 objects. Further cuts are then applied to define a \textit{gold} sample where the fraction of vote for each object belonging to any one of three morphological classes - ellipticals, spirals and point sources/artifacts, is always greater than 0.8. This gold sample contains $\sim$315,000 objects and is essentially equivalent to the clean sample of \citet{Lintott:zoo}. The neural network is run on the gold sample as well as the entire sample. 

It is also the case that faint disky objects are more likely to be classified as ellipticals unless the spiral arms can be clearly seen. The elliptical sample therefore also probably contains a reasonable number of lenticular systems and we therefore refer to this morphological class as early types throughout this paper. We therefore also consider a sample of objects with $r<17$ that is defined as our bright sample and should suffer from a lower level of contamination in the early type class. This magnitude limit is the same as that imposed on objects that were used to determine user weights in \citet{Lintott:zoo}. The bright sample contains $\sim$340,000 objects and has fewer ''well classified'' early types than the gold sample as discussed later. 

\begin{figure}
\begin{center}
\includegraphics[width=8.5cm,angle=0]{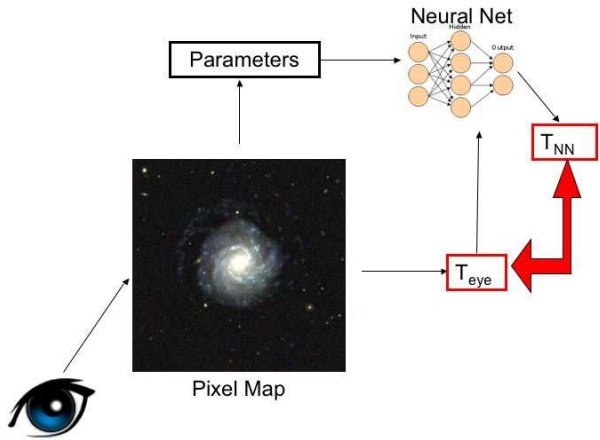}
\caption{Cartoon schematic of how both the human eye as well as machine learning algorithms such as artificial neural networks perform morphological classification and determine parameters such as those listed in Table \ref{tab:par1} and \ref{tab:par2} from the galaxy images.}
\label{fig:cartoon}
\end{center}
\end{figure}

\section{Artificial Neural Networks}

\label{sec:ann}

We use an artificial neural network code \citep{Ripley:81, Ripley:88, Bishop:NN, Lahav:95, Naim:95, Collister:ANNz} for classification in this paper. It has already been shown that, on a de Vaucouleurs type system, $T$, which spans values from -5 to 10, human expert classifiers agree to rms $\Delta_T=1.8$, and that such agreement can be obtained by a neural network when trained on the classifications of one of the experts \citep{Lahav:95, Naim:95}. The neural network used in our study is made up of several layers, each consisting of a number of nodes.  The first layer receives the input parameters described in detail in $\S$ \ref{sec:par} and the last layer outputs the probabilities for the object belonging to the three morphological classes. All nodes in the hidden layers in between are interconnected and connections between nodes $i$ and $j$ have an associated weight, $w_{ij}$.  A training set is used to minimise the cost function, $E$ (Eq. \ref{eq:cf}) with respect to the free parameters $w_{ij}$: 

\begin{equation}
E=\sum_{k}(T_{NN}(w_{ij},p_{k})-T_{eye,k})^2
\label{eq:cf}
\end{equation}

\noindent where $T_{NN}$ is the neural network probability of the object belonging to a particular morphological type, $p_k$ are the input parameters to the network and $T_{eye,k}$ are the fractional weighted votes in the training set in this case assigned by Galaxy Zoo users. 

If the data is noisy or the network is very flexible, a validation set may be used in addition to the training set to prevent over-fitting. During the initial setup, one has to specify the architecture of the neural network - the number of hidden layers and nodes in each hidden layer. We choose a neural network with two hidden layers with 2N nodes each, where N is the number of input parameters. The architecture of the network is therefore N:2N:2N:3.  Note that increasing the number of nodes further either by adding nodes to existing hidden layers or adding more hidden layers to the network, does not result in any substantial improvement to the classifications. The three nodes in the output layer give the probability of the galaxy being an early type, spiral and point source/artifact respectively between 0 and 1. Neural nets with this type of output are statistical Bayesian estimators and therefore the sum of all three outputs is roughly although rarely exactly equal to 1 (e.g. \citet{Lahav:96}; Appendix C and references therein). Note that this differs from the Galaxy Zoo fractional votes which always add up to exactly 1 over all four morphological classes - early types, spirals, point sources/artifacts and mergers. As mentioned earlier, the mergers are not classified by the neural network in this paper.   

\section{Input Parameters}

\label{sec:par}

When using an automated machine learning algorithm, the choice of input parameters may be crucial in determining how well the network can perform morphological classifications. Ideally, one wishes to choose a set of parameters that show marked differences across the three morphological classes. In addition, it may be useful to define a set of parameters that is independent of the distance to the object. For example, colours may be used instead of magnitudes and ratios of radii could replace individual radius estimates. Figure \ref{fig:cartoon} illustrates the key role that the input parameters play in the morphological classification. The human eye sees an image and performs morphological classifications. The same image is also used to derive input parameters such as those that will be discussed in this section. The neural network uses these parameters, as well as a training set based on the human classifications to derive its own morphological classifications, $T_{NN}$. These are then compared to the human classifications, $T_{eye}$ in order to assess the success of the machine-learning algorithm in reproducing what is seen by the human eye. 

In this section, we consider two sets of input parameters based on these criteria but make no additional effort to fine-tune and optimise these parameters to perform this morphological classification. The first set of parameters listed in Table \ref{tab:par1} have been used extensively in the literature for morphological classification in the SDSS e.g. \citep{Ball:04}. They include the (g-r) and (r-i) colours derived from the dereddened model magnitudes although these have not been k-corrected to the rest-frame. We do not use the k-corrected colours as this would require a redshift to be measured for the object and therefore reduce the total number of objects in our sample. Furthermore, the vast majority of objects in future large-scale photometric surveys will not have secure spectroscopic redshifts and we aim to assess how effective machine-learning is as a tool for morphological classification in such surveys. The other parameters considered are the axis ratios and log likelihoods associated with both a de Vaucouleurs and exponential fit to the two-dimensional galaxy image.The de Vaucouleurs profile is commonly used to describe the variation in surface brightness of an elliptical galaxy as a function of radius whereas the exponential profile is used to describe the disk component of a spiral galaxy. In addition, the log likelihood of the object being well fit by a PSF, \texttt{lnLstar}, helps in distinguishing extended galaxies from more point-like sources. Note that the sample only contains those objects that are well-fit by a PSF that also have spectra that are best-fit by a galaxy template. Stars have already been removed from the sample. The (g-r) and (r-i) colours have been chosen as the images used in the Galaxy Zoo classifications were composites of images in these three bands. All other parameters correspond to the i-band images only.

\begin{figure*}
\begin{center}
\begin{minipage}[c]{1.00\textwidth}
\centering
\includegraphics[width=8.5cm,height=7cm, angle=0]{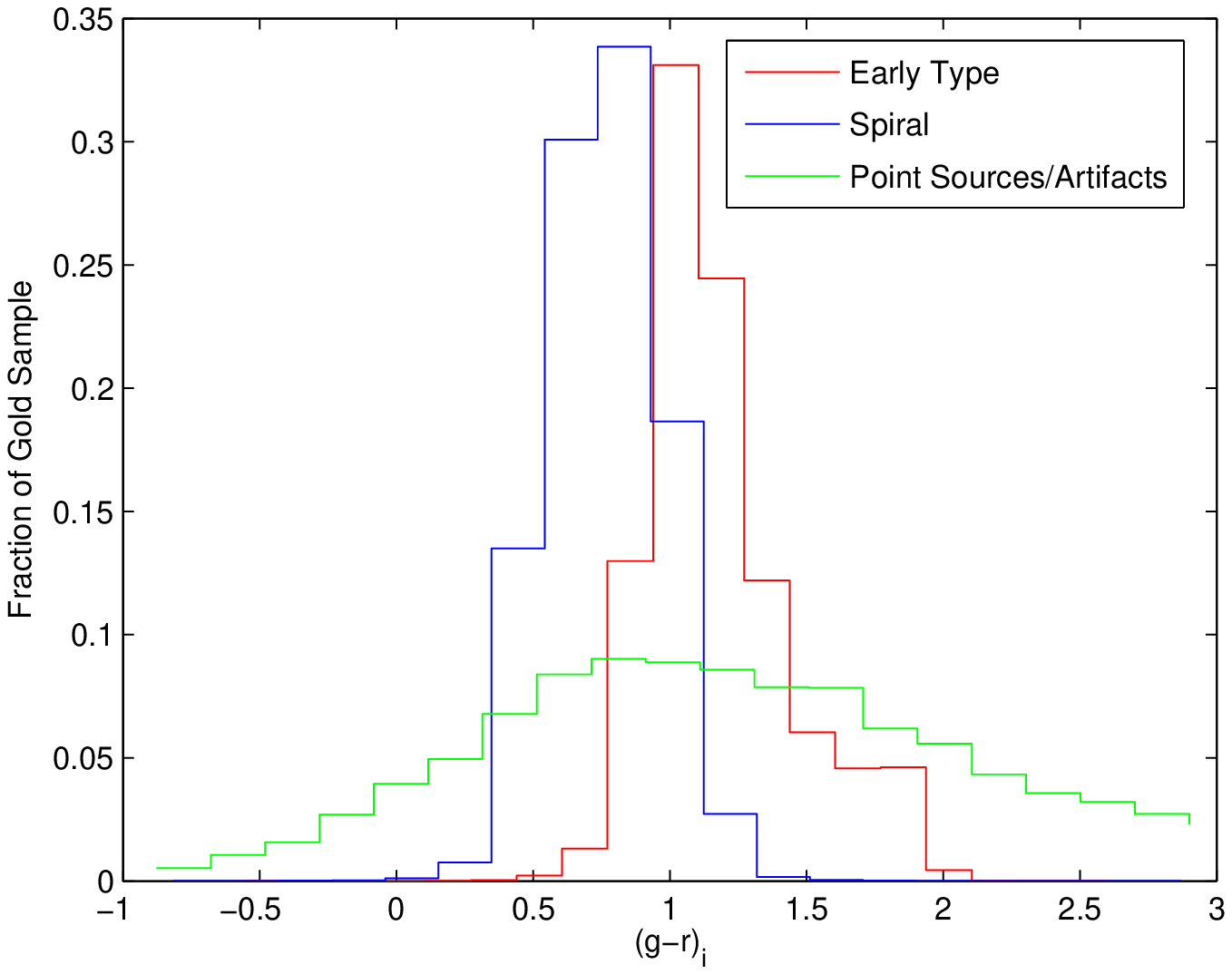}
\includegraphics[width=8.5cm,height=7cm, angle=0]{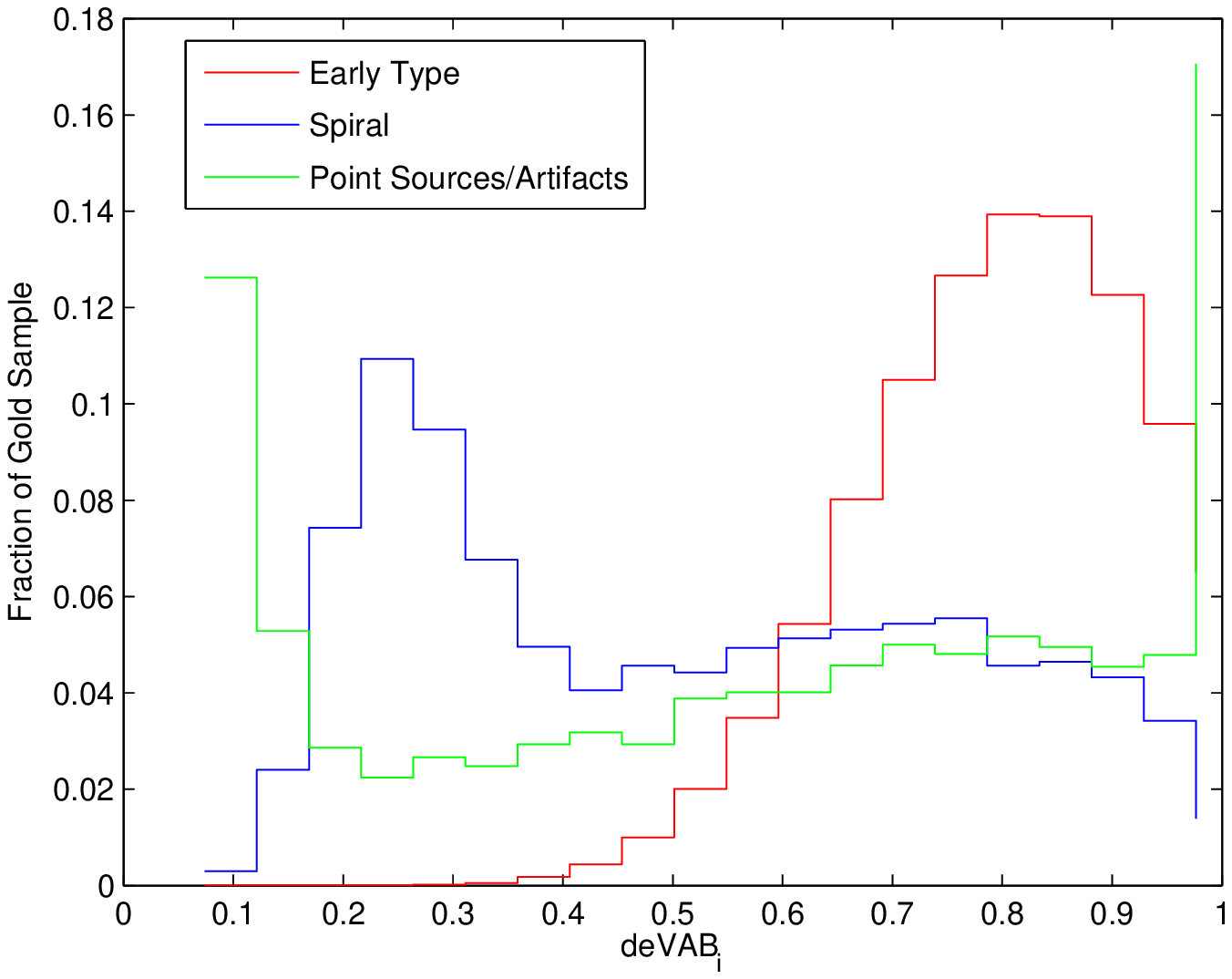}
\includegraphics[width=8.5cm,height=7cm, angle=0]{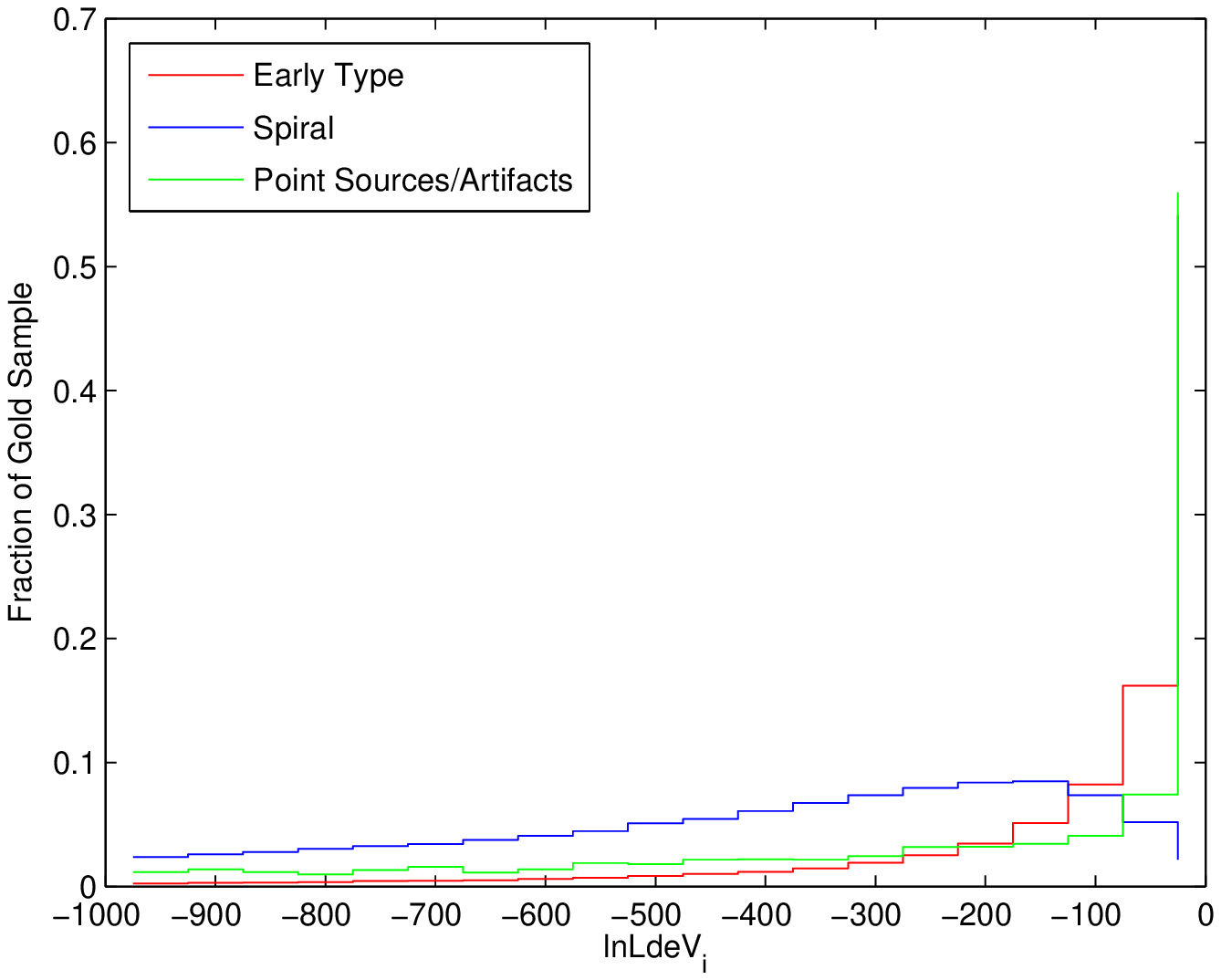}
\includegraphics[width=8.5cm,height=7cm, angle=0]{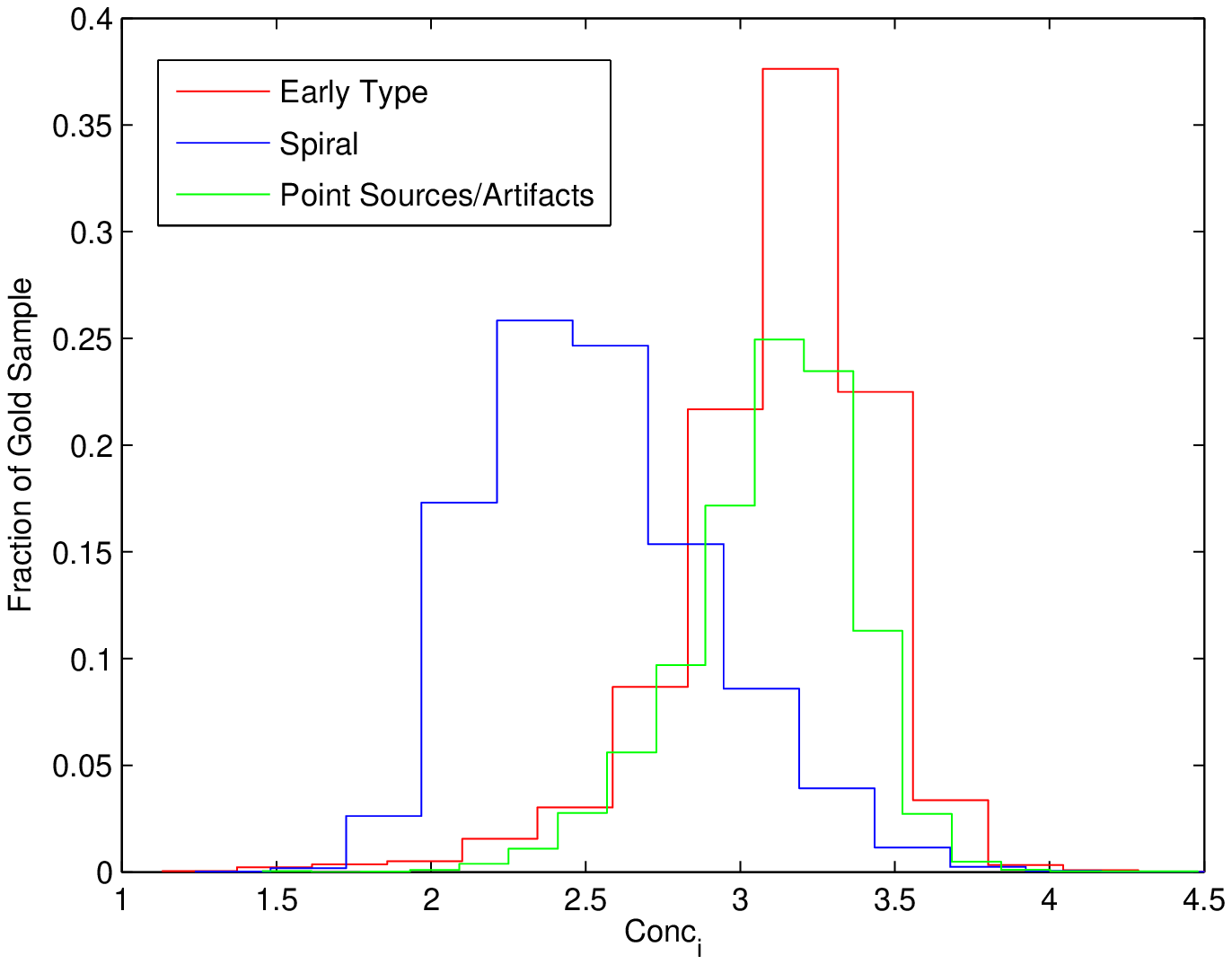}
\includegraphics[width=8.5cm,angle=0]{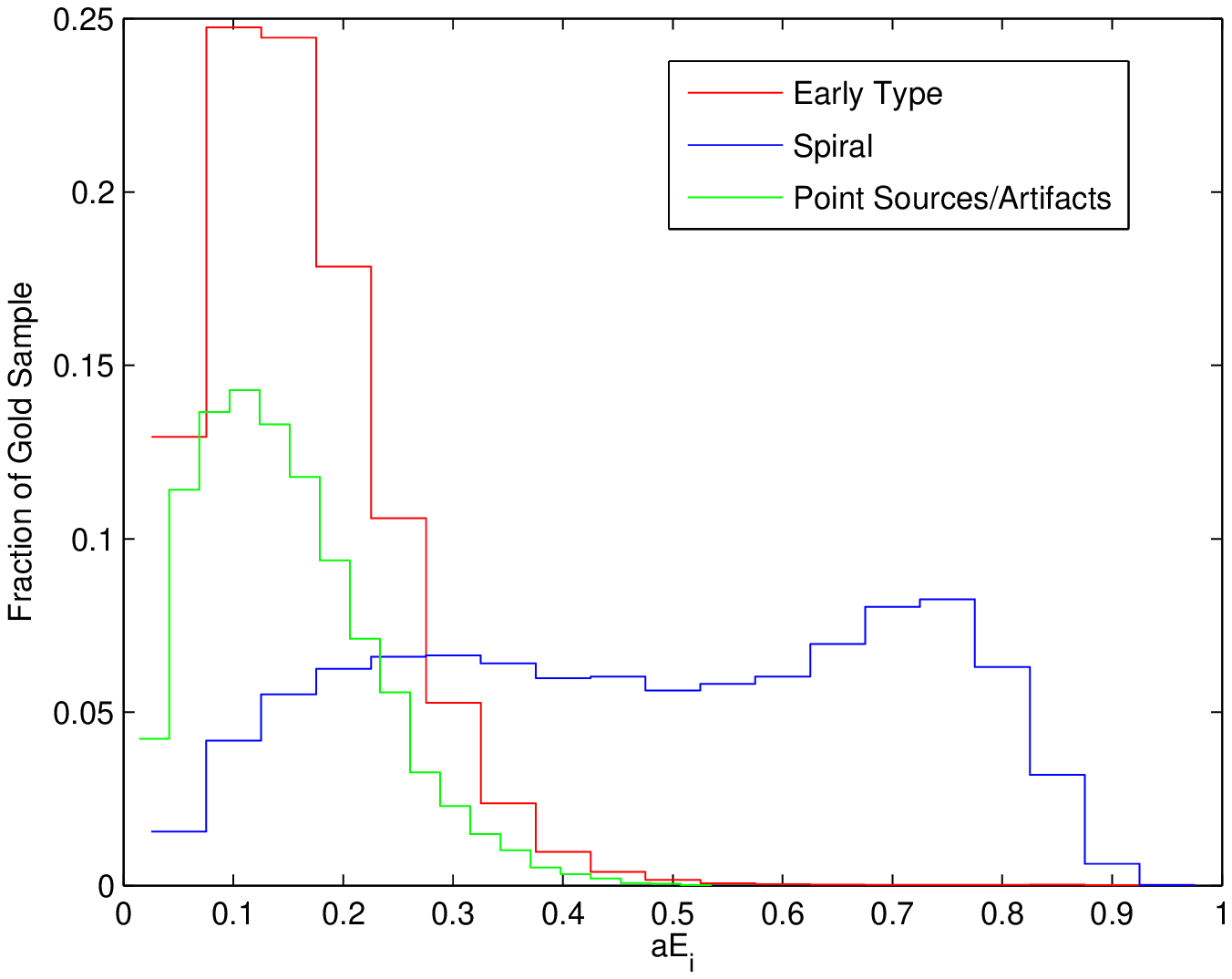}
\includegraphics[width=8.5cm,angle=0]{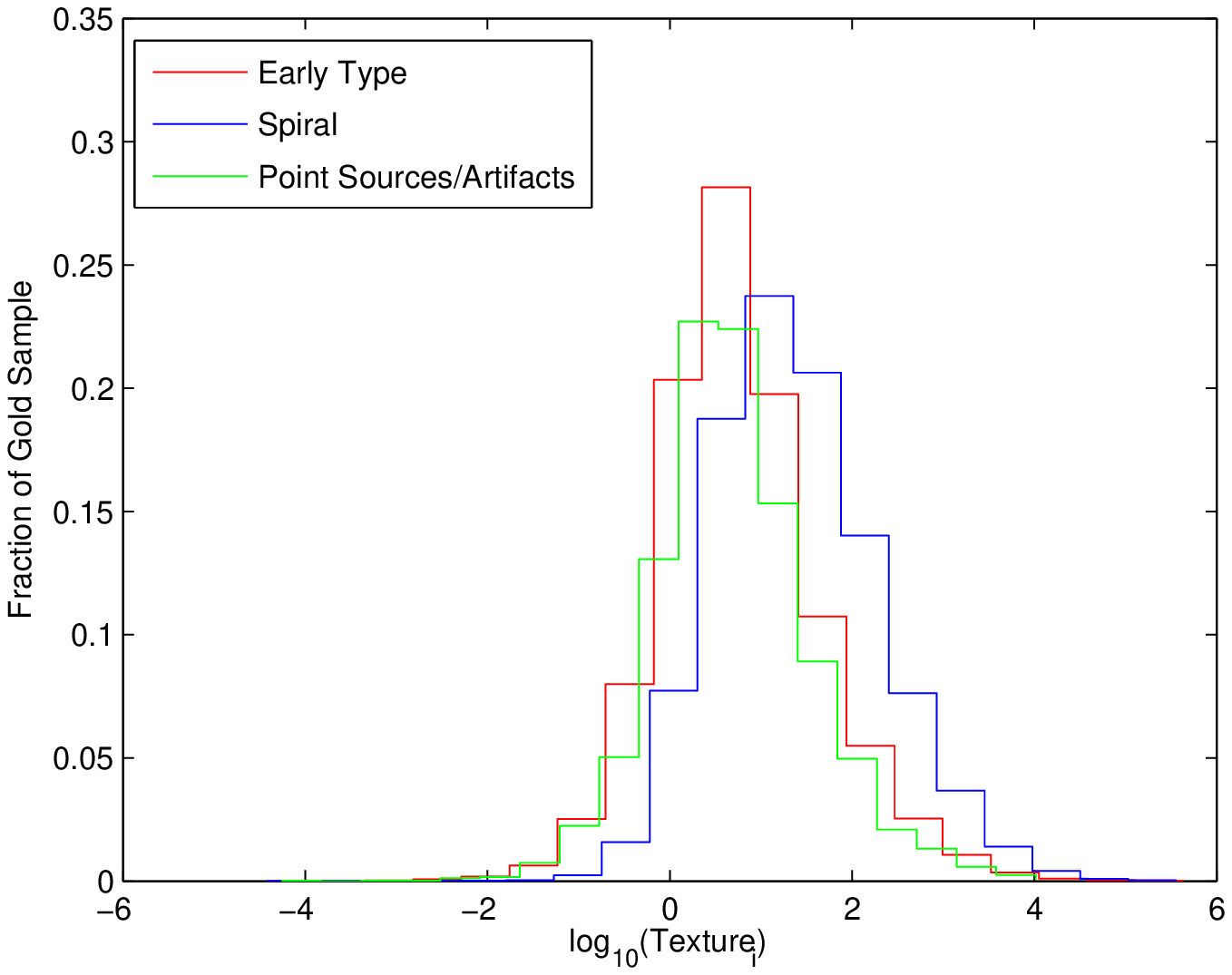}
\end{minipage}
\caption{Histograms of some of the input parameters in Tables \ref{tab:par1} and \ref{tab:par2} for the gold sample of early types, spirals and point sources/artifacts.}
\label{fig:hist1}
\end{center}
\end{figure*}

The distribution of the $(g-r)$ colour, de Vaucouleurs fit axis ratio and log likelihood fit to a de Vaucouleurs profile for the different morphological classes, is illustrated in Figure \ref{fig:hist1} where we plot the fraction of \textit{gold} early types, spirals and point sources/artifacts as a function of these parameters. These histograms are constructed using only objects with a fraction of vote greater than 0.8 in each of the three classes from Galaxy Zoo i.e. the \textit{gold} sample. This threshold is arbitrary and certainly results in the loss of many true early types and spirals from our sample. However, choosing a high threshold also ensures that the samples we use to construct histograms don't suffer much from contamination. Throughout the rest of this paper, the Galaxy Zoo early types, spirals and point source/artifacts always refer to the \textit{gold} sample with a fraction of vote greater than 0.8. Note that the fraction of vote is different from the classification probability although the two are highly correlated. Many objects with fractional votes less than 0.8 are also well classified early types or spirals but in this paper we only choose to consider those objects that are the most cleanly classified by members of the public for checking against the corresponding neural network classifications.  

We can immediately see from Figure \ref{fig:hist1} that different parameters allow us to distinguish between different morphological classes. As expected, early types are found to be redder than spirals whereas the point sources and artifacts have a wide range of colours. The axis ratio obtained from a de Vaucouleurs fit to the galaxy images is closer to unity for early type systems (typically $\sim$0.8) compared to spirals (typically $\sim$0.3) and has a bimodal distribution for the point sources and artifacts. The log likelihood associated with the de Vaucouleurs fit is also larger for the early types than the spirals and largest for the point sources and artifacts. 

\label{sec:params}

\begin{table}
  \begin{center}	
    \begin{tabular}{|l|c|}
      \hline
      Name & Description \\
      \hline
      \texttt{dered\_g-dered\_r} & (g-r) colour \\
      \texttt{dered\_r-dered\_i} & (r-i) colour \\
      \texttt{deVAB\_i} & de Vaucouleurs fit axis ratio \\
      \texttt{expAB\_i} & Exponential fit axis ratio \\
      \texttt{lnLexp\_i} & Exponential disk fit log likelihood \\
      \texttt{lnLdeV\_i} & de Vaucouleurs fit log likelihood \\
      \texttt{lnLstar\_i} & Star log likelihood \\
      \hline
    \end{tabular}	\vspace{2mm}
  \end{center}
  \caption{First Set of Input Parameters based on colours and profile fitting	\label{tab:par1}}
\end{table}

The second set of input parameters described in Table \ref{tab:par2} do not use the colours of galaxies or any parameters associated with profile fitting for morphological classification. Instead a new set of shape and texture parameters are used as well as the concentration. The concentration is given by the ratios of radii containing 90\% and 50\% of the Petrosian flux in a given band. \texttt{mRrCc} is the second moment of the object intensity in the CCD row and column directions measured using a scheme designed to have optimal signal-to-noise ratio \citep{Bernstein:02}.

\begin{equation}
\texttt{mRrCc}=<y^2>+<x^2>
\label{eq:mrrcc}
\end{equation}

\noindent where $x$ and $y$ are image coordinates relative to the centre of the object in question and, for example, 

\begin{equation}
<y^2>=\frac{\sum I(y,x)w(y,x)y^2}{\sum I(y,x)w(y,x)}
\label{eq:c2}
\end{equation}

 Moments are measured using a radial gaussian weight function, $w(y,x)$ iteratively adapted to the shape and size of the object. The ellipticity components, \texttt{mE1} and \texttt{mE2} defined in Eq \ref{eq:me1} and Eq \ref{eq:me2} and a fourth order moment (Eq \ref{eq:mcr4}) are also specified. 

\begin{equation}
\texttt{mE1}=\frac{<y^2>-<x^2>}{\texttt{mRrCc}}
\label{eq:me1}
\end{equation}

\begin{equation}
\texttt{mE2}=2\frac{<xy>}{\texttt{mRrCc}}
\label{eq:me2}
\end{equation}

\begin{equation}
\texttt{mCr4}=\frac{<(y^2+x^2)^2>}{\sigma^4}
\label{eq:mcr4}
\end{equation}

\noindent where $\sigma$ is the size of the gaussian weight.  We use the ellipticity components to define the adaptive ellipticity, \texttt{aE} for input into the neural network where:

\begin{equation}
\texttt{aE}=1-\sqrt{\frac{1-\sqrt{\texttt{mE1}^2+\texttt{mE2}^2}}{1+\sqrt{\texttt{mE1}^2+\texttt{mE2}^2}}}
\label{eq:ae}
\end{equation}

The final parameter in Table \ref{tab:par2} is the texture or coarseness parameter described in \citet{Yamauchi:05}. This essentially measures the ratio of the range of fluctuations in the surface brightness of the object to the full dynamic range of the surface brightness and is expected to vanish for a smooth profile but become non-zero if structures like spiral arms appear. 

The distribution of the concentration, adaptive ellipticity and texture parameters constructed using only the \textit{gold} objects in the Galaxy Zoo catalogue with a fraction of vote greater than 0.8, are shown in Figure \ref{fig:hist1}. The concentration parameter is larger for early types compared to spirals. This is consistent with the previous studies by \citet{Shimasaku:01} and \citet{Strateva:01} who find the inverse concentration index to be larger for spirals than for ellipticals. The adaptive ellipticity is large for the spirals, small for the early types and slightly smaller still for the point sources and artifacts. The texture parameter although roughly similar for the three morphological classes, is still slightly larger for spirals as compared to the early types and point sources suggesting that the latter have the smoothest surface brightness profiles. 

\begin{table}
  \begin{center}	
    \begin{tabular}{|l|c|}
      \hline
      Name & Description \\
      \hline
      \texttt{petroR90\_i/petroR50\_i} & Concentration \\
      \texttt{mRrCc\_i} & Adaptive (+) shape measure \\
      \texttt{aE\_i} & Adaptive Ellipticity \\
      \texttt{mCr4\_i} & Adaptive fourth moment \\
      \texttt{texture\_i} & Texture parameter \\
      \hline
    \end{tabular}	\vspace{2mm}
  \end{center}
  \caption{Second Set of Input Parameters based on adaptive moments	\label{tab:par2}}
\end{table}

We summarise the results of running the neural network with these different choices of input parameters in the next section.

\section{Results}

\label{sec:results}

The neural network code is run using three different sets of input parameters - (i) colours and profile-fitting parameters from Table \ref{tab:par1} (ii) concentration, adaptive shape parameters and texture from Table \ref{tab:par2} and (iii) a combined set of twelve parameters. We also define three samples on which the neural network is run. The first sample is the entire catalogue of $\sim$800,000 objects out of which 50,000 are used for training and 25,000 for validation. This is found to be a sufficiently large training set for these purposes and no significant improvement in the classifications is seen on increasing the size of the training set further. The second sample defined as the gold sample contains only objects with a fraction of vote greater than 0.8 assigned to them in any one of the three morphological classes by Galaxy Zoo users. This is essentially the same as the clean sample in \citet{Lintott:zoo}. This gold sample contains $\sim$315,000 objects and once again 50,000 are used for training and 25,000 for validation. In this gold sample, $\sim$65\% are early types, $\sim$30\% are spirals and $\sim$5\% are point sources and artifacts. These fractions should not however be interpreted as the true ratio of different morphological types in the Universe as the fraction of early types is much higher in a flux-limited sample such as this. In addition, the choice of an arbitrary fractional vote threshold of 0.8 for morphological classifications in Galaxy Zoo, means many true spirals and ellipticals do not make our cut. We also do not attempt to remove any classification bias as was done in \citet{Bamford:08} as this is not particularly important for the aims of this paper. Finally, we consider a bright sample with $r<17$ only. In this bright sample, $\sim$55\% are early types, $\sim$40\% are spirals and $\sim$5\% are point sources and artifacts and we can therefore see that the bias towards early types is somewhat reduced on removing the faint galaxies from the sample. 

\subsection{The Entire Sample}

\begin{figure*}
\begin{center}
\begin{minipage}[c]{1.00\textwidth}
\centering
\includegraphics[width=8.5cm,angle=0]{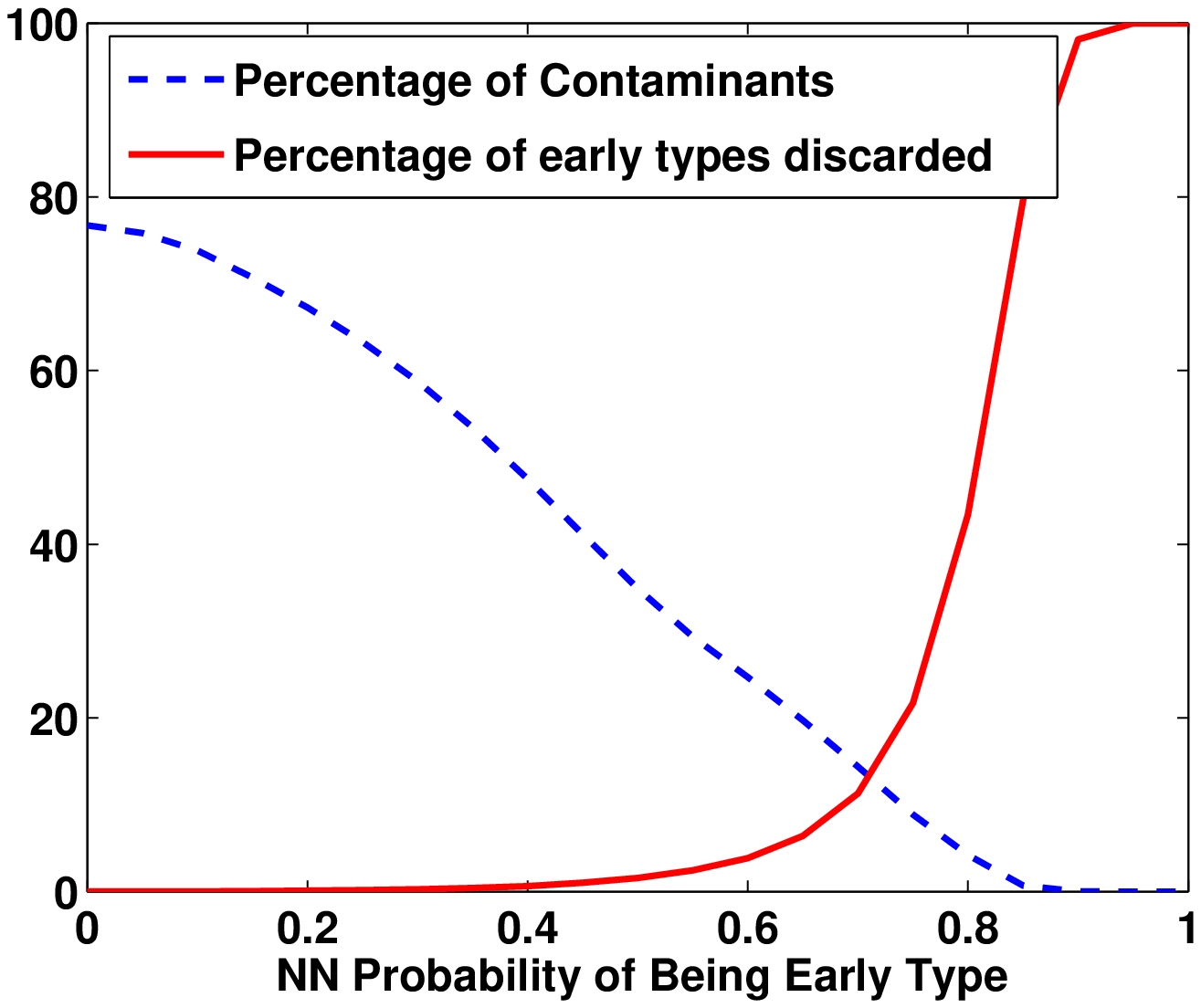}
\includegraphics[width=8.5cm,angle=0]{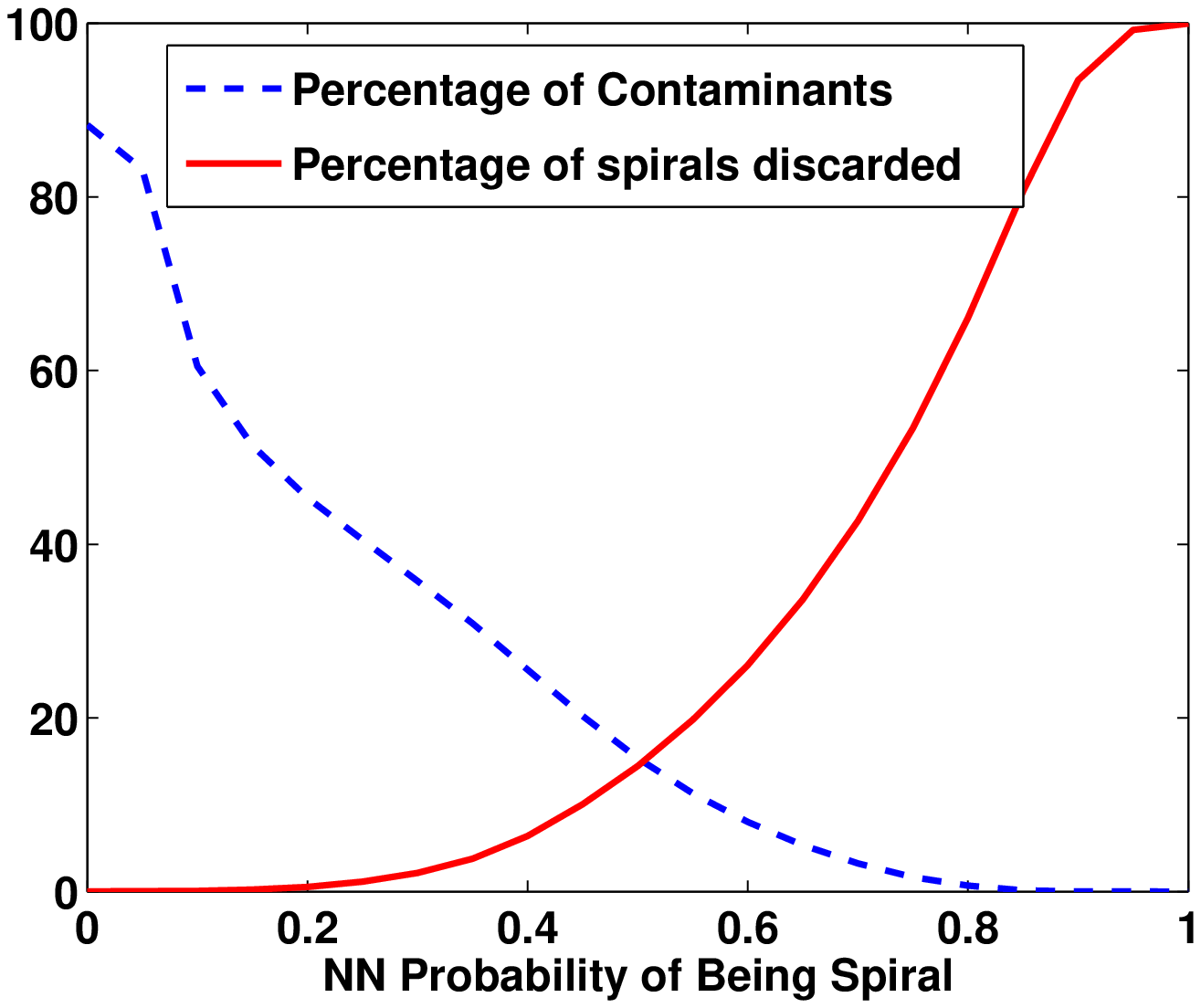}
\includegraphics[width=8.5cm,angle=0]{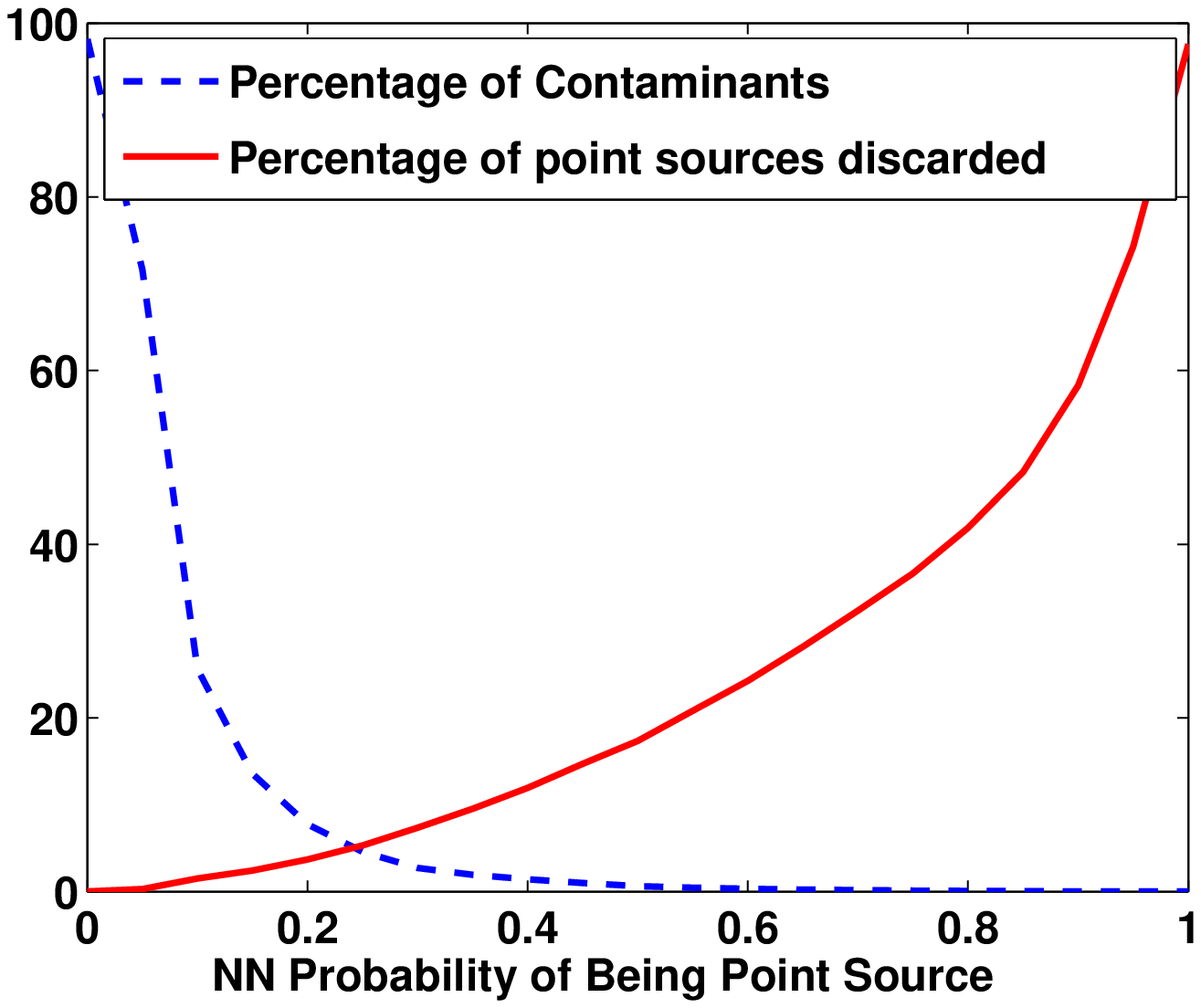}
\end{minipage}
\caption{The neural network probability of a galaxy being an early type (top left), spiral (top right) and point source/artifact (bottom) versus the percentage of contaminants as well as the percentage of Galaxy Zoo objects in these classes that are discarded. These results are obtained using the 7 input parameters in Table \ref{tab:par1} based on colours and traditional profile-fitting.}
\label{fig:rand1}
\end{center}
\end{figure*}

\begin{figure*}
\begin{center}
\begin{minipage}[c]{1.00\textwidth}
\centering
\includegraphics[width=8.5cm,angle=0]{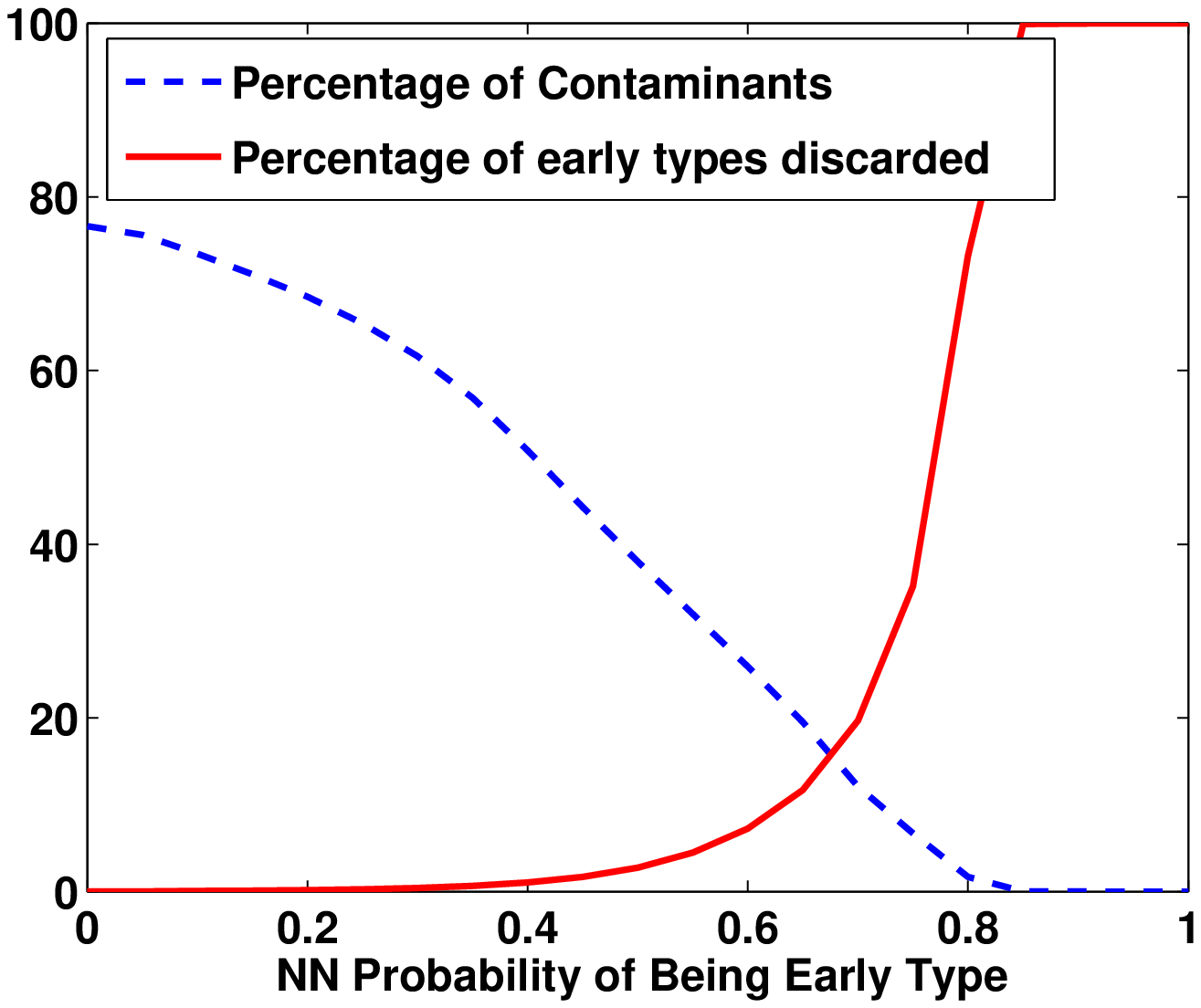}
\includegraphics[width=8.5cm,angle=0]{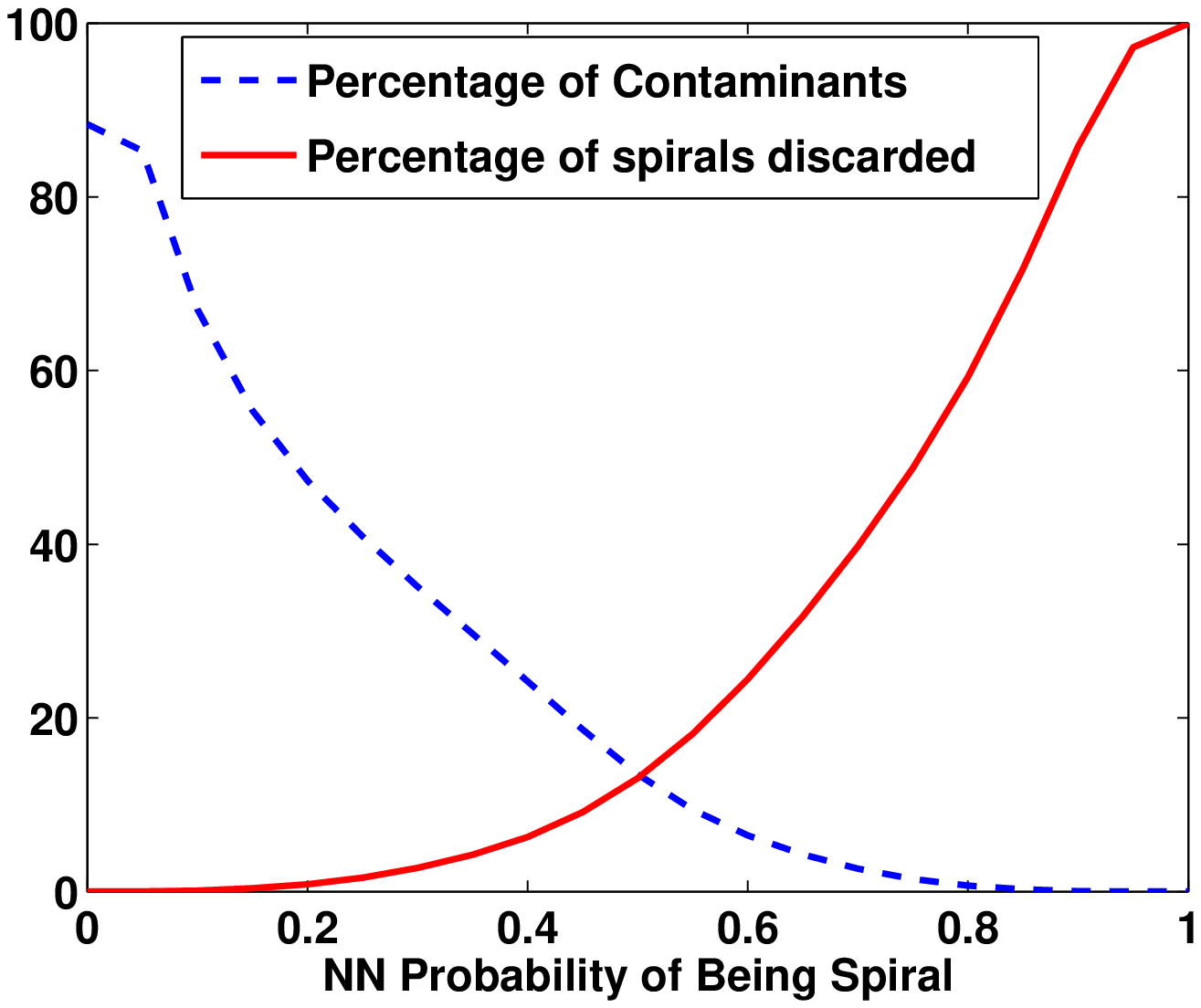}
\includegraphics[width=8.5cm,angle=0]{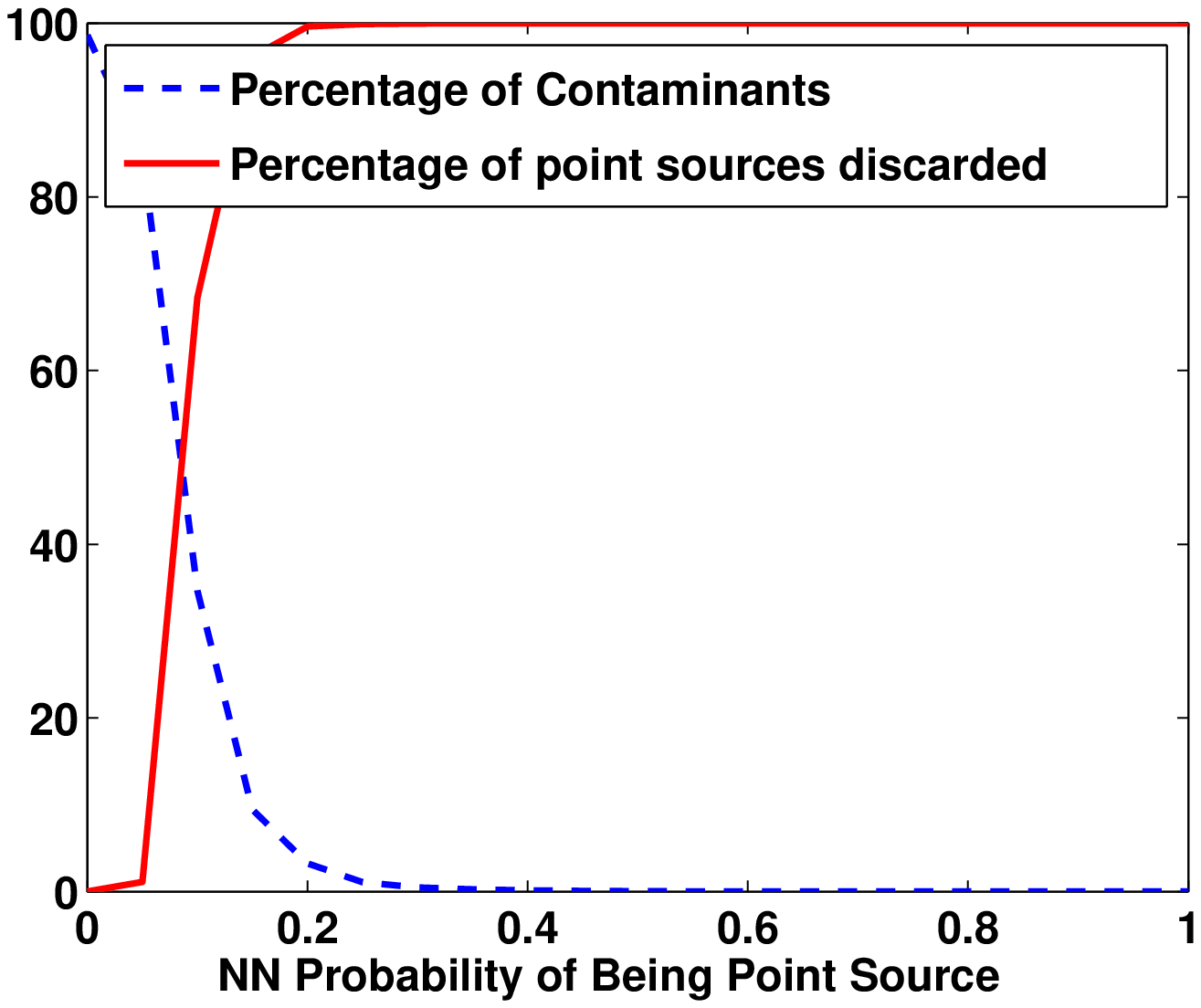}
\end{minipage}
\caption{The neural network probability of a galaxy being an early type (top left), spiral (top right) and point source/artifact (bottom) versus the percentage of contaminants as well as the percentage of Galaxy Zoo objects in these classes that are discarded. These results are obtained using the 5 input parameters in Table \ref{tab:par2} that use adaptive moments, concentration and texture.}
\label{fig:rand2}
\end{center}
\end{figure*}

\begin{figure*}
\begin{center}
\begin{minipage}[c]{1.00\textwidth}
\centering
\includegraphics[width=8.5cm,angle=0]{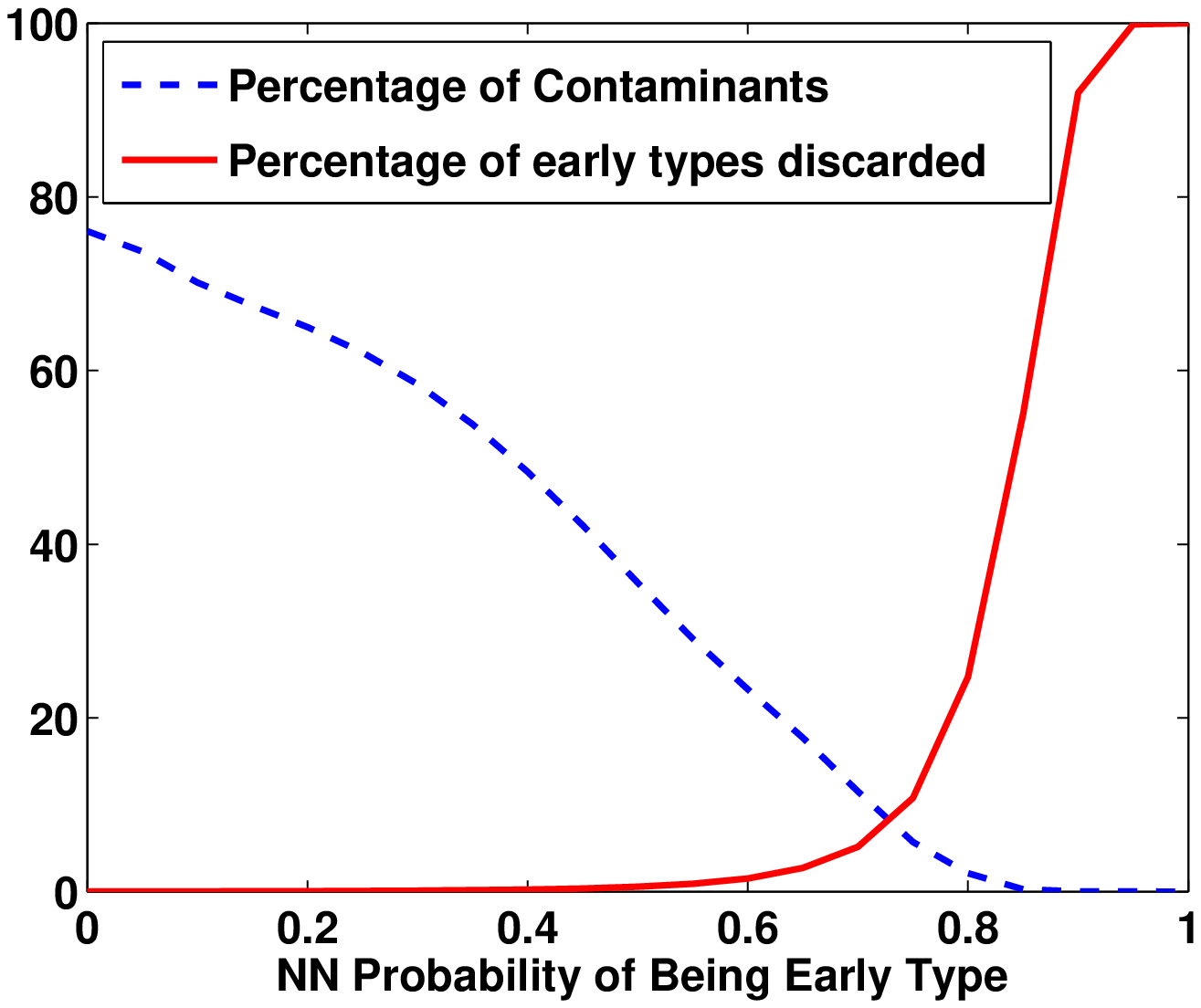}
\includegraphics[width=8.5cm,angle=0]{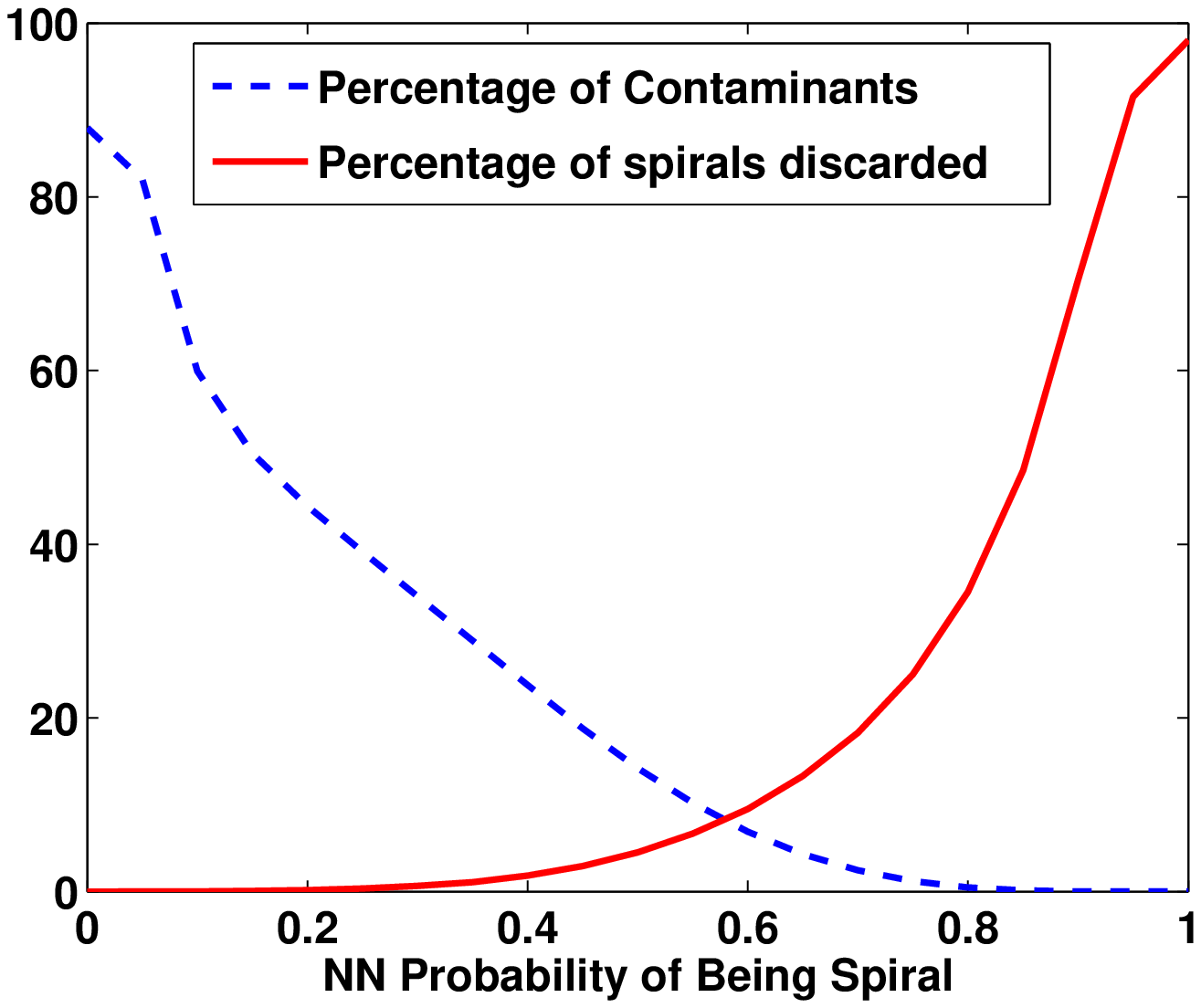}
\includegraphics[width=8.5cm,angle=0]{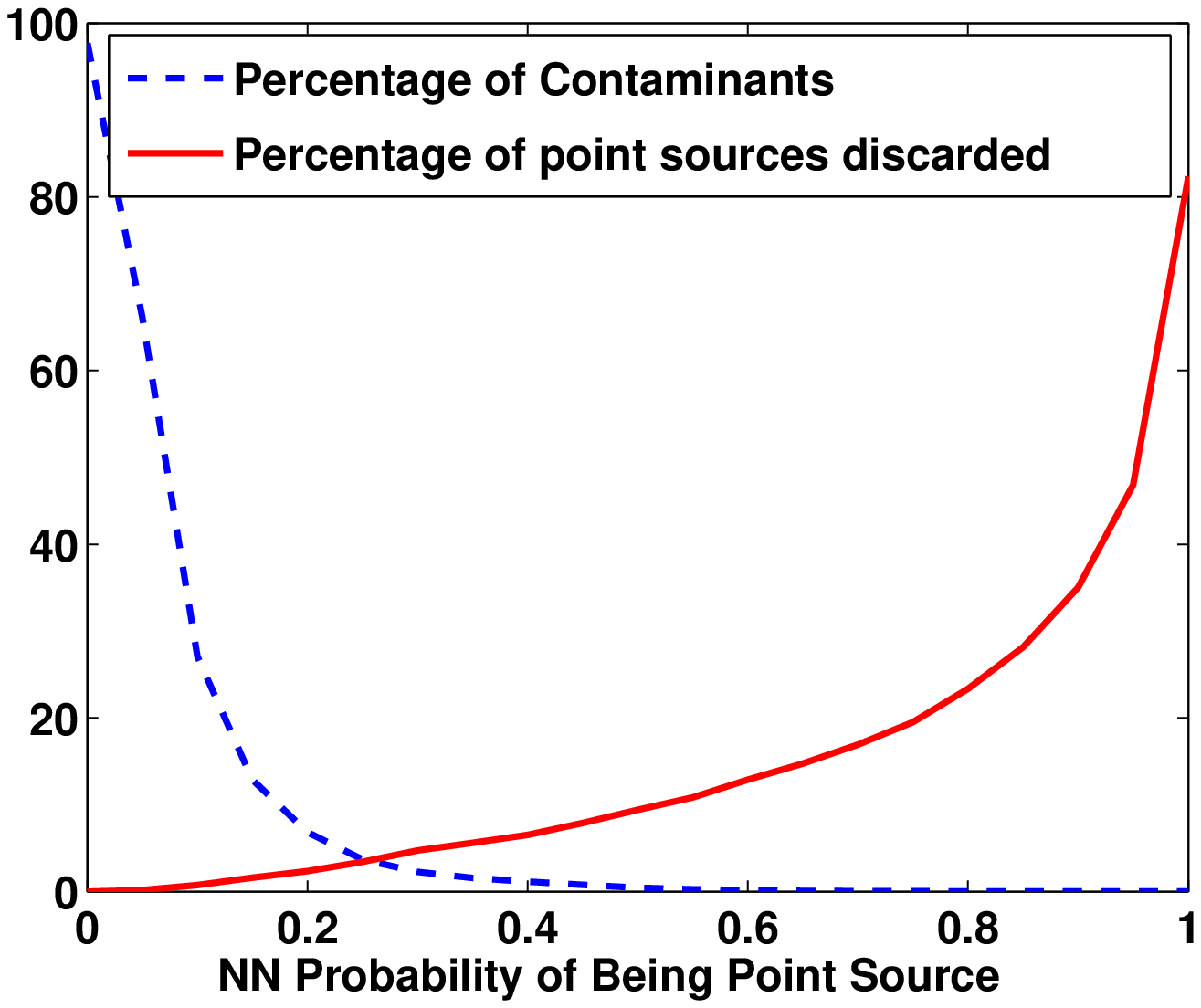}
\end{minipage}
\caption{The neural network probability of a galaxy being an early type (top), spiral (middle) and point source/artifact (bottom) versus the percentage of contaminants as well as the percentage of Galaxy Zoo objects in these classes that are discarded. These results are obtained using the combined set of 12 input parameters from Tables \ref{tab:par1} and \ref{tab:par2}.}
\label{fig:rand3}
\end{center}
\end{figure*}

In this section we summarise the results of running the neural network on the entire sample of objects using the three different sets of input parameters. In Figures \ref{fig:rand1}, \ref{fig:rand2} and \ref{fig:rand3}, we plot the NN probability of the object belonging to a morphological class versus the percentage of genuine objects in that class that are discarded on applying this probability threshold as well as percentage of contaminants that enter the sample. This allows us to determine the optimum probability threshold for the neural network that should be chosen for membership into each morphological class. This threshold is such that the number of contaminants equals the number of genuine objects in that class that are discarded. This is found to be 0.71 for early types, 0.50 for spirals, 0.24 for point sources/artifacts when using the parameters in Table \ref{tab:par1}; 0.68 for early types, 0.50 for spirals, 0.10 for point sources/artifacts when using the parameters in Table \ref{tab:par2}; and 0.73 for early types, 0.58 for spirals and 0.26 for point source/artifact when using the combined set of parameters. 

For the Galaxy Zoo classifications, we consider the fractional vote threshold to be 0.8 for the galaxy belonging to a particular morphological class. In Tables \ref{tab:rand1}, \ref{tab:rand2} and \ref{tab:rand3}, we summarise the results of running the neural net on the entire sample using the three different sets of input parameters. The tables give the percentage of early types, spirals and point sources/artifacts in Galaxy Zoo that are put into the different classes by the neural network after assuming the optimum NN probabilities already mentioned in each of the three classes. Throughout this paper, the lowercase names -  Early Type, Spiral, Point Source/Artifact - correspond to the Galaxy Zoo classifications and the upper case names - EARLY TYPE, SPIRAL, POINT SOURCE/ARTIFACT - correspond to the neural network classifications. Note that as we are attempting to minimise both the number of contaminants as well as the number of genuine objects discarded in each of the three classes, the total number of objects is not necessarily conserved. There may also be a significant number of poorly classified objects with a spread of likelihoods over two or more morphological classes. These poorly classified objects will have a probability of less than the chosen optimum threshold probability in all three morphological classes. Some of these objects that are poorly classified by the neural network may have been well classified by Galaxy Zoo which is why the sum of the columns in Table \ref{tab:rand1} are typically less than 100\%. Some objects may also be put into more than one class due to the assumed neural network threshold probabilities. For example, an object with a neural net probability of greater than 0.1 in the POINT SOURCE/ARTIFACT class, could potentially, also have a neural net probability of greater than 0.5 in the SPIRAL class and therefore be classified as both. This is why the sum of the columns in Table \ref{tab:rand2} are typically greater than 100\%. In the tables presented throughout this analysis, we are not summarising the results of unique classifications for every single object but rather the best results that could be obtained for each subsample assuming neural network threshold probabilities that result in the same number of contaminants into the sample as genuine objects that are discarded. If we were to consider the best results for the entire sample as a whole, the chosen probability thresholds would probably be different.      

\begin{table*} 
 \begin{center}	
    \begin{tabular}{cc|c|c|c}
      & & \multicolumn{3}{c}{\textbf{GALAXY ZOO}} \\
      & & Early Type & Spiral & Point Source/Artifact \\
      \hline
      \hline
      \textbf{A} & EARLY TYPE & 87\% & 0.3\% & 0.3\% \\
      \textbf{N} & SPIRAL & 0.6\% & 86\% & 2.2\% \\
      \textbf{N} & POINT SOURCE/ARTIFACT & 0.7\% & 0.5\% & 95\% \\
      \hline
\end{tabular}	\vspace{2mm}
  \end{center}
  \caption{Summary of results for entire sample when using input parameters specified in Table \ref{tab:par1} - colours and traditional profile-fitting. \label{tab:rand1}}
\end{table*} 

 \begin{table*} 
 \begin{center}	
    \begin{tabular}{cc|c|c|c}
      & & \multicolumn{3}{c}{\textbf{GALAXY ZOO}} \\
      & & Early Type & Spiral & Point Source/Artifact \\
      \hline
      \hline
      \textbf{A} & EARLY TYPE & 84\% & 0.5\% & 85\% \\
      \textbf{N} & SPIRAL & 1\% & 87\% & 0.8\% \\
      \textbf{N} & POINT SOURCE/ARTIFACT & 32\% & 6.5\% & 32\% \\
      \hline
\end{tabular}	\vspace{2mm}
  \end{center}
  \caption{Summary of results for entire sample when using input parameters specified in Table \ref{tab:par2} - adaptive moments. \label{tab:rand2}}
\end{table*} 

 \begin{table*} 
 \begin{center}	
    \begin{tabular}{cc|c|c|c}
      & & \multicolumn{3}{c}{\textbf{GALAXY ZOO}} \\
      & & Early Type & Spiral & Point Source/Artifact \\
      \hline
      \hline
      \textbf{A} & EARLY TYPE & 92\% & 0.07\% & 0.6\% \\
      \textbf{N} & SPIRAL & 0.1\% & 92\% & 0.08\% \\
      \textbf{N} & POINT SOURCE/ARTIFACT & 0.2\% & 0.2\% & 96\% \\
      \hline
\end{tabular}	\vspace{2mm}
  \end{center}
  \caption{Summary of results for entire sample when using input parameters specified in Table \ref{tab:par1} and Table \ref{tab:par2} \label{tab:rand3}}
\end{table*} 

Using the traditional colour and profile-fitting parameters, 87\% of early types, 86\% of spirals and 95\% of point sources/artifacts are correctly classified by the neural network. We note however that while colours are sensitive to the star-formation history of a galaxy, the morphology essentially measures the dynamic history. Although the two are correlated, they are not necessarily the same and therefore it is important to use colours in conjunction with other parameters when performing morphological classifications. Looking more closely at the Galaxy Zoo early types that were misclassified by the neural net as spirals as well as the spirals that are misclassified by the neural net as early types, there is some evidence that these galaxies may be red spirals or blue ellipticals. 45 Galaxy Zoo early types are classified as spirals with a probability of greater than 0.8 by the neural net. Similarly, 40 Galaxy Zoo spirals are classified as early types with a probability greater than 0.8. Out of these, 21 early types and 9 spirals have SDSS spectra available. Using the criterion in \citet{Baldry:04} to isolate red and blue galaxies, we find that 6 out of the 9 spirals are red and 10 out of the 21 early types are blue. In other words, the colour information used as an input to the neural network may be biasing the morphological classification. However, due to the small numbers of misclassified galaxies with SDSS spectra available, a definite statement cannot be made and we leave this to future work. 

The adaptive shape parameters are very good for distinguishing between spirals and early types and these parameters result in accurate classifications for 84\% of early types and 87\% of spirals. However, this set of parameters gives very poor results for point source/artifact and only 28\% are correctly classified by the neural network. This is because, these parameters are very similar for point sources/artifacts and early types as can be seen in some of the histograms in Figure \ref{fig:hist1}. The spirals on the other hand have very different adaptive shape parameters. As there are many more early types in the training set compared to point sources/artifacts, the neural network cannot differentiate between the two and assigns most of the point sources/artifacts to be early types. Also Figure \ref{fig:rand2} shows that the optimum neural network probability for point source/artifact classification using this set of input parameters is as low as 0.1. This means that many objects will be classified both as a point source and a galaxy assuming this probability and for this reason, the sums of the columns in Table \ref{tab:rand2} are always greater than 100\% as there are objects in common between the classes. We also investigate whether the evidence for a bias due to colour in the morphological classifications is removed once the colours are removed as input parameters to the network. With the adaptive shape parameters as inputs we find 208 Galaxy Zoo early types to be misclassified as spirals with a probability greater than 0.8 by the neural network and similarly, 26 spirals are misclassified as early types with the same probability. Out of these, 116 early types and only 8 spirals have SDSS spectra. 2 out of the 8 spirals are red but 46 of the 116 early types are blue once again suggesting that there may still remain a bias, especially against blue ellipticals even when the colour parameters are removed as inputs to the neural net. This may suggest that atleast some blue ellipticals have more structure than their red counterparts. However, we emphasise that the sample sizes here are too small to make any quantitative statement about the extent of this bias and we defer this to future work. 

On adding the profile-fitting and colour parameters to the adaptive shape parameters, the results are now considerably improved for all three classes. 92\% of early types, 92\% of spirals and 96\% of point sources/artifacts are correctly classified by the neural network. \citet{Lintott:zoo} have compared the Galaxy Zoo classifications to those by professional astronomers and find an agreement of better than 97\% between these samples. However, the samples used in \citet{Lintott:zoo} are the MOSES sample of \citet{Schawinski:MOSES} whose objective was to generate a very clean set of elliptical galaxies, and the detailed classifications of \citet{Fukugita:07} which are very sensitive to the E/Sa boundary. If the professional astronomers were set the same task as the Galaxy Zoo users - i.e. a clean division of spirals and ellipticals in SDSS -  the scatter between them and the Galaxy Zoo users may well be worse than this. We have shown that with a set of twelve intelligently chosen parameters, that are easily available but still by no means optimal for performing morphological classification, the neural network results agree to better than 90\% with those from Galaxy Zoo. It can therefore certainly be expected that by using a set of better tuned input parameters to the network, the machine learning algorithm will be able to classify galaxies with comparable or less scatter than that produced by human classifications. 

\subsection{The Gold Sample}

We now describe the results of running the neural network on our gold sample using the three different sets of input parameters. These results are summarised in Tables \ref{tab:clean1}, \ref{tab:clean2} and \ref{tab:clean3} where we consider the percentage of Galaxy Zoo early types, spirals and point sources/artifacts that have also been put into these classes by the neural network assuming a probability of greater than 0.8 now for the neural network classification. Once again it can be seen that the second set of input parameters involving just the concentration, texture and the adaptive shape parameters, performs very poorly in classifying the point sources and artifacts. However, the combined set of input parameters leads to correct neural network classifications for 97\% of early types and 97\% of spirals, on par with the agreement between Galaxy Zoo and professional astronomers. The success for point sources/artifacts is somewhat lowered compared to using the entire sample. This is because when considering the entire sample we assume that all objects with a neural network point source/artifact probability of greater than $\sim$0.2 do indeed belong to the point sources/artifacts class whereas in this section we require the probability to be greater than 0.8.    

 \begin{table*} 
 \begin{center}	
    \begin{tabular}{cc|c|c|c}
      & & \multicolumn{3}{c}{\textbf{GALAXY ZOO}} \\
      & & Early Type & Spiral & Point Source/Artifact \\
      \hline
      \hline
      \textbf{A} & EARLY TYPE & 93\% & 0.6\% & 0.9\% \\
      \textbf{N} & SPIRAL & 0.7\% & 90\% & 2.3\% \\
      \textbf{N} & POINT SOURCE/ARTIFACT & 0.04\% & 0.07\% & 82\% \\
      \hline
\end{tabular}	\vspace{2mm}
  \end{center}
  \caption{Summary of results for gold sample when using input parameters specified in Table \ref{tab:par1} - colours and traditional profile-fitting. \label{tab:clean1}}
\end{table*} 

 \begin{table*} 
 \begin{center}	
    \begin{tabular}{cc|c|c|c}
      & & \multicolumn{3}{c}{\textbf{GALAXY ZOO}} \\
      & & Early Type & Spiral & Point Source/Artifact \\
      \hline
      \hline
      \textbf{A} & EARLY TYPE & 92\% & 0.8\% & 92\% \\
      \textbf{N} & SPIRAL & 0.6\% & 89\% & 0.5\% \\
      \textbf{N} & POINT SOURCE/ARTIFACT & 0\% & 0\% & 0\% \\
      \hline
\end{tabular}	\vspace{2mm}
  \end{center}
  \caption{Summary of results for gold sample when using input parameters specified in Table \ref{tab:par2} - adaptive moments. \label{tab:clean2}}
\end{table*} 

 \begin{table*} 
 \begin{center}	
    \begin{tabular}{cc|c|c|c}
      & & \multicolumn{3}{c}{\textbf{GALAXY ZOO}} \\
      & & Early Type & Spiral & Point Source/Artifact \\
      \hline
      \hline
      \textbf{A} & EARLY TYPE & 97\% & 0.2\% & 1.3\% \\
      \textbf{N} & SPIRAL & 0.1\% & 97\% & 0.2\% \\
      \textbf{N} & POINT SOURCE/ARTIFACT & 0.05\% & 0.02\% & 86\% \\
      \hline
\end{tabular}	\vspace{2mm}
  \end{center}
  \caption{Summary of results for gold sample when using input parameters specified in Table \ref{tab:par1} and Table \ref{tab:par2} \label{tab:clean3}}
\end{table*} 

\subsection{The Bright Sample}

In this section, we run the neural network code using the combined set of twelve input parameters on our bright sample with $r<17$. This allows us to perform two tests. Firstly, we look at whether the bright galaxies in general have better classifications compared to the entire sample. This is done by comparing Tables \ref{tab:rand3} and \ref{tab:bright1}. Secondly, we train our neural network on the bright sample and use this trained network to perform morphological classifications for the entire sample. This allows us to quantify the effects of magnitude incompleteness in the training set on the morphological classifications. The results are summarised in Tables \ref{tab:bright1} and \ref{tab:bright2}. Once again the probability threshold required in the neural net output for classification is determined by requiring the percentage of contaminants to be equal to the percentage of genuine objects in that class that are discarded on applying the threshold. This optimum probability threshold is very similar for the bright sample to those shown in Figure \ref{fig:rand3} but slightly higher in all three classes when the bright galaxies are used for training and used to classify all galaxies. 

 \begin{table*} 
 \begin{center}	
    \begin{tabular}{cc|c|c|c}
      & & \multicolumn{3}{c}{\textbf{GALAXY ZOO}} \\
      & & Early Type & Spiral & Point Source/Artifact \\
      \hline
      \hline
      \textbf{A} & EARLY TYPE & 94\% & 0.1\% & 0.3\% \\
      \textbf{N} & SPIRAL & 0.1\% & 92\% & 0.1\% \\
      \textbf{N} & POINT SOURCE/ARTIFACT & 0.2\% & 0.2\% & 98\% \\
      \hline
\end{tabular}	\vspace{2mm}
  \end{center}
  \caption{Summary of results for bright sample when using input parameters specified in Table \ref{tab:par1} and Table \ref{tab:par2} \label{tab:bright1}}
\end{table*} 

 \begin{table*} 
 \begin{center}	
    \begin{tabular}{cc|c|c|c}
      & & \multicolumn{3}{c}{\textbf{GALAXY ZOO}} \\
      & & Early Type & Spiral & Point Source/Artifact \\
      \hline
      \hline
      \textbf{A} & EARLY TYPE & 91\% & 0.05\% & 0.8\% \\
      \textbf{N} & SPIRAL & 0.08\% & 92\% & 0.1\% \\
      \textbf{N} & POINT SOURCE/ARTIFACT & 2\% & 0.1\% & 95\% \\
      \hline
\end{tabular}	\vspace{2mm}
  \end{center}
  \caption{Summary of results for entire sample when using input parameters specified in Table \ref{tab:par1} and Table \ref{tab:par2} and only bright galaxies with $r<17$ to train the network.\label{tab:bright2}}
\end{table*} 

Comparing Tables \ref{tab:rand3} and \ref{tab:bright1}, we can see that the results are slightly better when performing morphological classsifications on the bright sample with $r<17$ compared to the entire sample with $r<17.77$, in both cases using a complete training set. This is to be expected as it is easier to distinguish between early types and spirals in a brighter sample. When the training is performed using an incomplete training set with $r<17$ and all objects with $r<17.77$ classified, the neural network still manages to perform these classifications with more than 90\% agreement with Galaxy Zoo users. The magnitude incompleteness in the training set doesn't seem to affect the classifications as most of the input parameters to the neural network considered in this study have been chosen to be distance independent and so their distribution doesn't really change from a shallow to a deeper sample. This is promising for using automated machine learning algorithms to perform morphological classsification for future deep surveys using the Galaxy Zoo classifications on the shallower SDSS survey as a training set. 

\section{Conclusions}

\label{sec:conclusion}

In this study, we have used a machine-learning algorithm based on artificial neural networks to perform morphological classifications for almost 1 million objects from the Sloan Digital Sky Survey that were classified by eye as part of the Galaxy Zoo project. The neural network is trained on 75000 objects and using a well defined set of input parameters that are also distance independent, we are able to reproduce the human classifications for the rest of the objects to better than 90\% in three morphological classes - early types, spirals and point sources/artifacts. The Galaxy Zoo catalogue provides us with a training set of unprecedented size for automated morphological classifications via machine learning. Specifically, we draw the following conclusions: 

\begin{itemize}

\item{Using colours and profile-fitting parameters as inputs to the neural network, 87\% of early type classifications, 86\% of spiral classifications and 95\% of point source/artifact classifications agree with those obtained by the human eye. However, there is some evidence to suggest that a non-negligible fraction of red spirals and blue ellipticals are misclassified by the network when using this set of parameters.}

\item{When parameters that rely on an adaptive weighted scheme for fitting the galaxy images are used, the neural network is unable to distinguish between early types and point source/artifact as these parameters are very similar for the two classes.}

\item{A combination of the profile fitting and adaptive weighted fitting parameters results in better than 90\% agreement between classifications by humans and those by the neural network for all three morphological classes. This is approaching the success of Galaxy Zoo users in reproducing the classifications by professional astronomers and will certainly surpass this with a more finely tuned set of input parameters than we have considered in this paper.}

\item{The optimum neural network probability for a galaxy belonging to a particular morphological class is such that the percentage of contaminants is equal to the percentage of genuine objects in that class that are discarded on cutting the sample using this threshold. This optimum probability depends both on the input parameters as well as the morphological class of the object. We find that early types generally have a high optimum probability ($\sim0.7$) whereas the point sources and artifacts have a very low optimum probability ($\sim$0.2). Therefore the same object could be put into more than one class by the neural network if the classifications were performed using the optimum threshold probabilities.}

\item{For our gold sample, the early type and spiral classifications by the neural network match those by the human eye to better than 95\%.} 

\item{Using a bright sample to train the neural network and performing morphological classifications for a deeper and fainter sample still results in better than 90\% agreement between the neural network and human classifications in all three morphological classes. This is because we have deliberately chosen our input parameters to be distance independent. }

\item{However, other sources of incompleteness in the training sets also need to be examined before the role of the Galaxy Zoo data in training morphological classifiers for future surveys, can be fully understood.}

\end{itemize}

The penultimate point in particular illustrates the power of the machine learning algorithm in fully exploiting data from future wide-field imaging surveys such as the Dark Energy Survey, Pan-STARRS, HyperSuprime-Cam, LSST, Euclid etc. Such surveys will obtain images for hundreds of millions of objects down to very deep magnitude limits. We have shown that by using the wealth of information made available through the Galaxy Zoo project as a training set, the machine learning algorithm can quickly and accurately classify the vast numbers of objects that will make up future data sets into early types, spirals and point sources/artifacts. However, if the Galaxy Zoo catalogue is to be used as a training set for automated machine learning classifications of mergers with the next generation of galaxy surveys, a more robust catalogue of visually classified mergers such as that of \citet{Darg:09} needs to be used. Also, it is worth emphasising that the images obtained from the next generation of wide-field surveys will need to have the necessary pixel size and resolution required to derive photometric parameters such as those used as inputs to the neural network in this paper.  

This paper has also examined the effect of magnitude incompleteness in the training set on automated morphological classifications. In the future, more work needs to be done on investigating other sources of incompleteness in the training set before this data can be used effectively to train machine-learning morphological classifiers for future surveys. For example, in this paper we have found some evidence that a non-negligible proportion of galaxies that are misclassified by the neural network are either red spirals or blue ellipticals. This misclassification is almost certainly due to the sparsity of such objects in the training set and needs to be addressed in future machine learning papers that use this data. We have also only used objects with a fraction of vote greater than 0.8 from Galaxy Zoo to compare to our neural network classifications. In future work, we hope to address how excluding ''intermediate'' objects from our comparison is likely to bias our results. Nevertheless, we conclude that morphological classification via machine-learning looks promising in allowing for many more detailed studies of the processes involved in galaxy formation and evolution with the next generation of galaxy surveys. 

\section*{Acknowledgements}

This work has depended on the participation of many members of the public in visually classifying SDSS galaxies on the Galaxy Zoo website. We thank them for their efforts. We would also like to thank the MNRAS anonymous referee whose comments and suggestions helped substantially improve this paper. MB and SB acknowledge support from STFC. OL and FBA acknowledge the support of the Royal Society via a Wolfson Royal Society Research Merit Award and a Royal Society URF respectively. CJL acknowledges support from The Leverhulme Trust and a STFC Science and Society Fellowship. FBA also acknowledges support from the Leverhulme Trust via an Early Careers Fellowship. Support of the work of KS was provided by NASA through Einstein Postdoctoral Fellowship grant number PF9-00069 issued by the Chandra X-ray Observatory Center, which is operated by the Smithsonian Astrophysical Observatory for and on behalf of NASA under contract NAS8-03060. 


\bibliography{}

\end{document}